\documentclass{aa}
\usepackage{graphicx}
\usepackage{amsfonts,amssymb,latexsym}
\pdfoutput=1

\def\ga{\,\,\raise0.14em\hbox{$>$}\kern-0.76em\lower0.28em\hbox
{$\sim$}\,\,}
\def\la{\,\,\raise0.14em\hbox{$<$}\kern-0.76em\lower0.28em\hbox
{$\sim$}\,\,}

\def\Msun{$M_{\odot}$}
\def\be{\begin{equation}} 
\def\ee{\end{equation}}
\def\beqy{\begin{eqnarray}}
\def\eeqy{\end{eqnarray}}
\def\bmlet{\begin{mathletters}}
\def\emlet{\end{mathletters}}

\begin{document}
\title{The decompression of the outer neutron star crust and r-process nucleosynthesis}

\author{S. Goriely \inst{1}, N. Chamel \inst{1}, H.-T. Janka \inst{2}, \and J.M. Pearson \inst{3}}

\offprints{S. Goriely}

\institute{
Institut d'Astronomie et d'Astrophysique, Universit\'e Libre de Bruxelles,  CP 226, B-1050 Brussels, Belgium
\and 
Max-Planck-Institut f\"ur Astrophysik, Postfach 1317, 85741 Garching, Germany
\and 
D\'epartement de Physique, Universit\'e de Montr\'eal, Montr\'eal (Qu\'ebec), H3C 3J7 Canada }

\date{Received --; accepted --}

\abstract
{The rapid neutron-capture process, or r-process, is known to be
fundamental  for explaining the origin of approximately half
of the $A>60$ stable nuclei observed in nature. In recent years nuclear
astrophysicists have developed more and more sophisticated r-process models, by
adding new astrophysical or nuclear physics ingredients to
explain the solar system composition in a satisfactory way. Despite these efforts, the astrophysical site of the r-process remains unidentified.}
{The composition of the neutron star outer crust  material is investigated after the decompression that follows its possible ejection. } 
{The composition of the outer crust of a neutron star is estimated before and after decompression. Two different possible initial conditions are considered, namely an idealized crust composed of cold catalyzed matter and a crust initially in nuclear statistical equilibrium at temperatures around $10^{10}$~K. } 
{We show that in this second case before decompression and at temperatures typically corresponding to  $8~10^9$~K, the Coulomb effect owing to the high densities in the crust leads to an overall composition of the outer crust in neutron-rich nuclei with a mass distribution close to the solar system r-abundance distribution. Such distributions differ, however, from the solar one due to a systematic shift in the second peak to lower values. After decompression, the capture of the few neutrons per seed nucleus available in the hot outer crust leads to a final distribution of stable neutron-rich nuclei with a mass distribution of $80 \le A  \le 140$ nuclei in excellent agreement with the solar distribution, provided the outer crust is initially at temperatures around $8~10^9$~K and all layers of the outer crust are ejected.}
{The decompression of the neutron star matter from the outer crust provides suitable conditions for a
robust r-processing of the light species, i.e., r-nuclei with $A \le 140$.  The final composition should carry the imprint of the temperature at which the nuclear statistical equilibrium is frozen prior to the ejection.}


\titlerunning{The r-process nucleosynthesis in neutron stars}
\authorrunning{S. Goriely et al.}

\maketitle



\maketitle

\section{Introduction}
\label{sect:intro}
The r-process, or the rapid neutron-capture process, of stellar nucleosynthesis
is invoked to explain the  production of the stable  (and some long-lived radioactive) neutron-rich nuclides 
that are heavier than iron and  observed in stars of various metallicities, as well as in the solar
system (for a review, see~Arnould et al. 2007). In recent years nuclear
astrophysicists have developed more and more sophisticated r-process models,
trying to explain the solar system composition in a satisfactory way by
adding new astrophysical or nuclear physics ingredients.
The r-process remains the most complex nucleosynthetic process for modelling from 
the astrophysics as well as nuclear-physics points of view. The site(s) of
the r-process has (have) not been identified yet, since all the proposed scenarios face serious
problems. Complex---and often exotic---sites have been considered in the hope of
identifying astrophysical conditions in which the production of neutrons is high
enough to give rise to a successful r-process. 

Progress in the modelling of type-II
supernovae and $\gamma$-ray bursts has raised a lot of excitement about the so-called
neutrino-driven wind environment. However, until now a successful r-process cannot be 
obtained {\it ab initio} without tuning the relevant parameters (neutron
excess, entropy, expansion timescale) in a way that is not supported by the most 
sophisticated existing models. 
Although these scenarios remain promising, especially in view of their potential to
significantly contribute to the galactic enrichment (Argast et al. 2004), they remain
handicapped by large uncertainties associated mainly with the still incompletely
understood mechanism that is responsible for the  supernova explosion and the persistent
difficulties obtaining suitable r-process conditions in self-consistent dynamical
explosion and neutron-star cooling models (H\"udepohl et al. 2010; Fischer et al.\ 2010). 
In addition, predictions of the detailed composition of the ejected matter remain 
difficult owing to the remarkable sensitivity of r-process nucleosynthesis to 
uncertainties of the ejecta properties.

Early in the development of the theory of nucleosynthesis, an alternative to the r-process 
in high-temperature supernova environments was proposed (Tsuruta et al. 1965).
It relies on the tendency of matter at high densities (typically $\rho > 10^{10}$ gcm$^{-3}$) to be composed of nuclei lying on the neutron-rich side of the valley of nuclear stability as a result of endothermic free-electron
 captures. This so-called `neutronization' of matter is possible even at zero temperature. 
 The astrophysical plausibility of this  scenario in accounting for the production of the r-nuclides has long been questioned.  It remained largely unexplored until the study  of the decompression of cold neutronized matter resulting from tidal effects of a black hole on a neutron star  (NS) companion (Lattimer et al. 1977; Meyer 1989). 
Recently, special attention has been paid to NS mergers following the 
confirmation by  hydrodynamic simulations that a non-negligible amount of matter can be
ejected (Janka et al. 1999; Rosswog et al. 2004; Oechslin et al.  2007).  
The ejection of initially cold, decompressed NS matter might also happen in other 
astrophysical scenarios like giant flares in soft-gamma repeaters, the explosion of an NS 
eroded below its minimum mass (e.g. Sumiyoshi et al.\ 1998),
or the equatorial shedding of material from very rapidly rotating supramassive or 
ultramassive NSs (see Arnould et al. 2007 for more details).

The composition of the NS matter of the inner crust, before and after decompression, has already been studied in detail in Goriely et al. (2005) and  Arnould et al. (2007). It was found that the final composition of the material ejected from the inner crust is expected to depend on the initial density,  at least for the upper part of the inner crust at $\rho_{\mathrm{drip}} \le \rho {\rm [g/cm^{3}]} \le 10^{12}$ (where $\rho_{\mathrm{drip}}\simeq
4.2~10^{11} {\rm g/cm^3}$ is the neutron drip density). For the deeper inner-crust layers ($\rho > 10^{12}~{\rm g/cm^{3}}$), large neutron-to-seed ratios  drive the nuclear flow into the very heavy mass region, leading to fission recycling.  As a consequence, the abundance distribution is now independent of the initial conditions, especially the initial density. In both cases, the abundance distribution was found to be in close agreement with the solar distribution for $A>130$ nuclei (Goriely et al. 2005; Arnould et al. 2007).

Although the outer crust is far less massive than the inner crust, the ejection
of the inner crust cannot take place without at the same time leading to the ejection of at least some outer crust material. The outer crust typically amounts to $10^{-5}$ to $10^{-4}$~\Msun, depending on the NS mass and radius (Pearson et al. 2011) so that depending on the frequency and fraction of its possible ejection, it may or may not contribute significantly to the galactic enrichment. We will not get into these considerations in the present paper, but restrict ourselves to estimating the composition of the outer crust material before (Sect.~\ref{sect_t}) as well as after ejection (Sect.~\ref{sect_d}), assuming that nuclear statistical equilibrium (NSE) could be established before ejection occurred. We shall consider  two cases: the first one corresponds to an initially cold NS crust and the second one to an NS  crust initially in NSE at a temperature of the order of $10^{10}$~K at which nuclear statistical equilibrium (NSE) is responsible for setting the composition prior to the ejection.

\section{Composition of the NS outer crust in equilibrium}
\label{sect_t}

The composition of the outer crust needs to be estimated in each layer, each one being characterized by a given density and pressure.  The overall abundance distribution for the outer crust ($0\le \rho(r)\le \rho_{\mathrm{drip}}$) can then be  integrated over a pressure column $P(r)$ (where $r$ is the radial coordinate) given by the Tolman-Oppenheimer-Volkoff (TOV) equations (considering non-rotating NS)
\beqy
\frac{dP}{dr} = -G\frac{\{\mathcal{M}(r)+ 4\pi r^3 P(r)/c^2\}\{\mathcal{E}(r) +
P(r)\}}{r\{rc^2 - 2G\mathcal{M}(r)\}}
\eeqy
with
\beqy
\frac{d\mathcal{M}}{dr} = 4\pi \mathcal{E}(r)r^2/c^2\quad ,
\eeqy
in which $\mathcal{E}$ denotes the total energy density, including rest mass (in the outer crust,  $\mathcal{E} \simeq \rho c^2$) and $\mathcal{M}$ is the gravitational mass.
We choose here a standard NS with a mass of 1.5~\Msun and 13~km radius. Since the equation of state (EoS) and the composition remain sensitive to the nuclear physics ingredients adopted (Pearson et al. 2011), we will consider here three different Hartree-Fock-Bogolyubov (HFB) mass models, namely the Skyrme-HFB mass models HFB-19 and HFB-21 (Goriely et al. 2010) and the Gogny-HFB mass model  D1M  (Goriely et al. 2009). For each mass model the EoS has been consistently taken into account in solving the TOV equation. In the three cases, a baryonic mass of about $5~10^{-5}$~\Msun (Pearson et al. 2011) is found in the outer crust of such an NS and about 90\% of the outer-crust mass is located in the density range of $6~10^{10}~{\rm g/cm^3} \la \rho \le \rho_{\mathrm{drip}}$. For densities above the drip density, free neutrons are present and their contributions to the EoS should be included. A transitional regime into the inner crust can be included in the calculation by estimating the pressure and free energy associated with the corresponding gas of free neutrons obtained self-consistently with the effective interactions BSk19, BSk21 or D1M corresponding to the above-mentioned mass tables.

\subsection{The cold NS crust}
\label{sec_t0}

A fairly common picture of an NS (Baym et al. 1971; Pethick \& Ravenhall 1995) is that they consist
of ``cold catalyzed matter", i.e., electrically neutral matter in its absolute 
ground state in the sense of complete nuclear and beta equilibrium at 
temperature $T = 0$. Detailed calculations of the composition of the NS material initially at a density below the drip density ($\rho_{\mathrm{drip}}=4.2 \times 10^{11}~{\rm g\,cm}^{-3}$) have been performed over the years (Baym et al. 1971; Pethick \& Ravenhall 1995; Haensel \& Pichon 1994; R\"uster et al. 2006; Pearson et al. 2011).
 We have repeated the calculation of Baym et al. (1971) with the updated nuclear physics inputs to determine the  outer-crust composition. This calculation minimizes the free Gibbs energy per nucleon at $T=0$. All details can be found in  Pearson et al. (2011) and will not be repeated here.

For densities above $3~10^9  {\rm g/cm^3}$, the outer crust is essentially made of $N=50$ and $N=82$ neutron-rich nuclei.  More precisely, for $10^9 \le \rho[{\rm g/cm^3}] \la 5~10^{10}$, we find $N=50$ nuclei with $80 \le A \le 86$. At these densities, only nuclei with experimentally known masses are involved. 
This is not the case at higher densities, i.e., in the most massive part of the outer crust, where we use the HFB-19, HFB-21 or D1M  mass tables to  complement experimental masses. For $10^{11} \le \rho[{\rm g/cm^3}] \le \rho_{\mathrm{drip}}$, $N=82$ nuclei with $120 \le A \le 126$ populate the outer crust.

\begin{figure}
\centering
\includegraphics[scale=0.30]{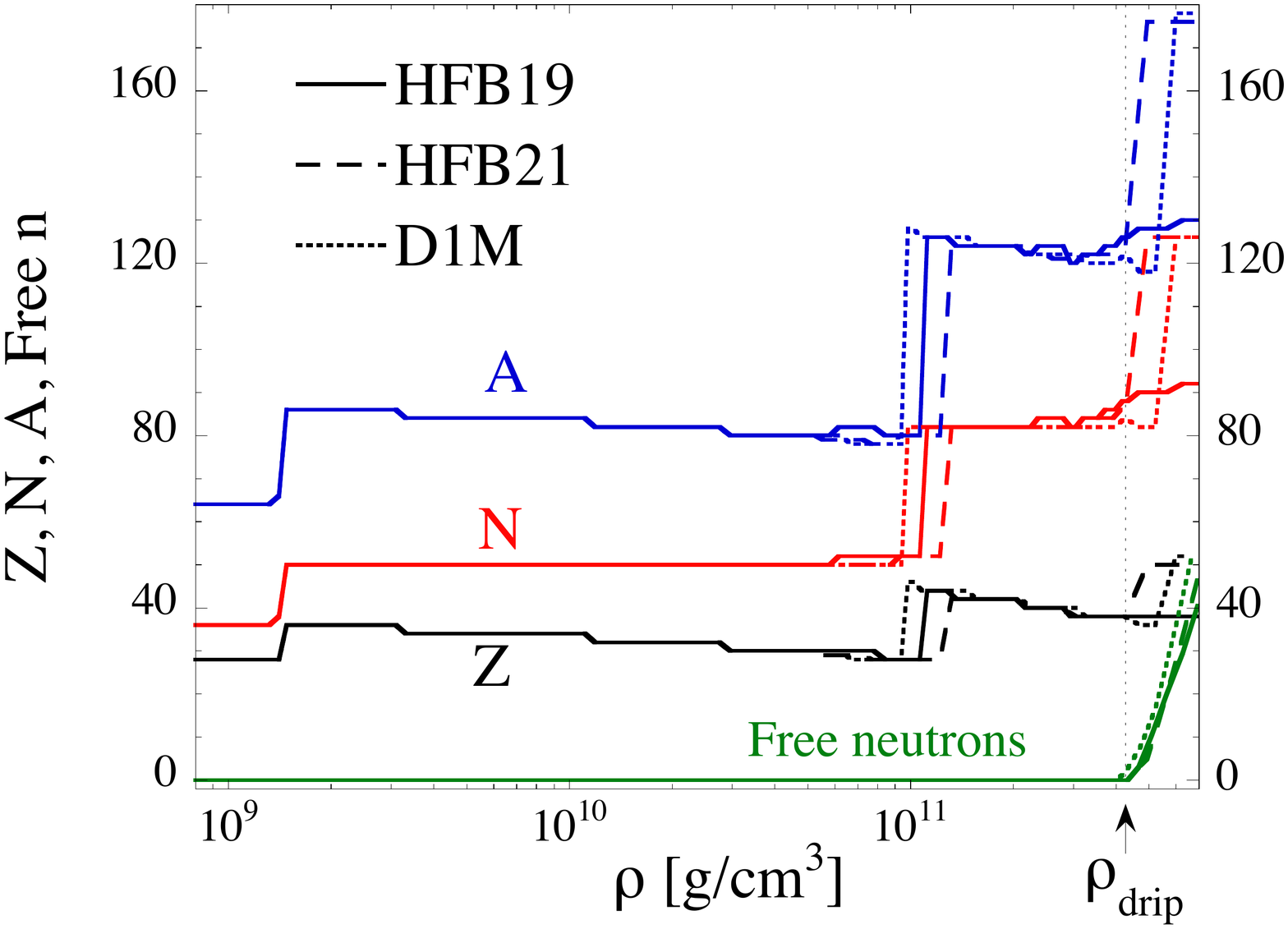}
\caption{Ground-state composition (charge, neutron and mass numbers as well as free neutrons) of the outer  crust and of the shallow layers of the inner crust as a function of the density. Predictions with HFB-19 masses  (solid lines) (Goriely et al. 2010) are compared with those obtained with the D1M masses (dotted lines) (Goriely et al. 2009). Experimental  masses (Audi et al. 2003, 2010) are used whenever available.}
\label{fig01}
\end{figure}

\begin{figure}
\centering
\includegraphics[scale=0.32]{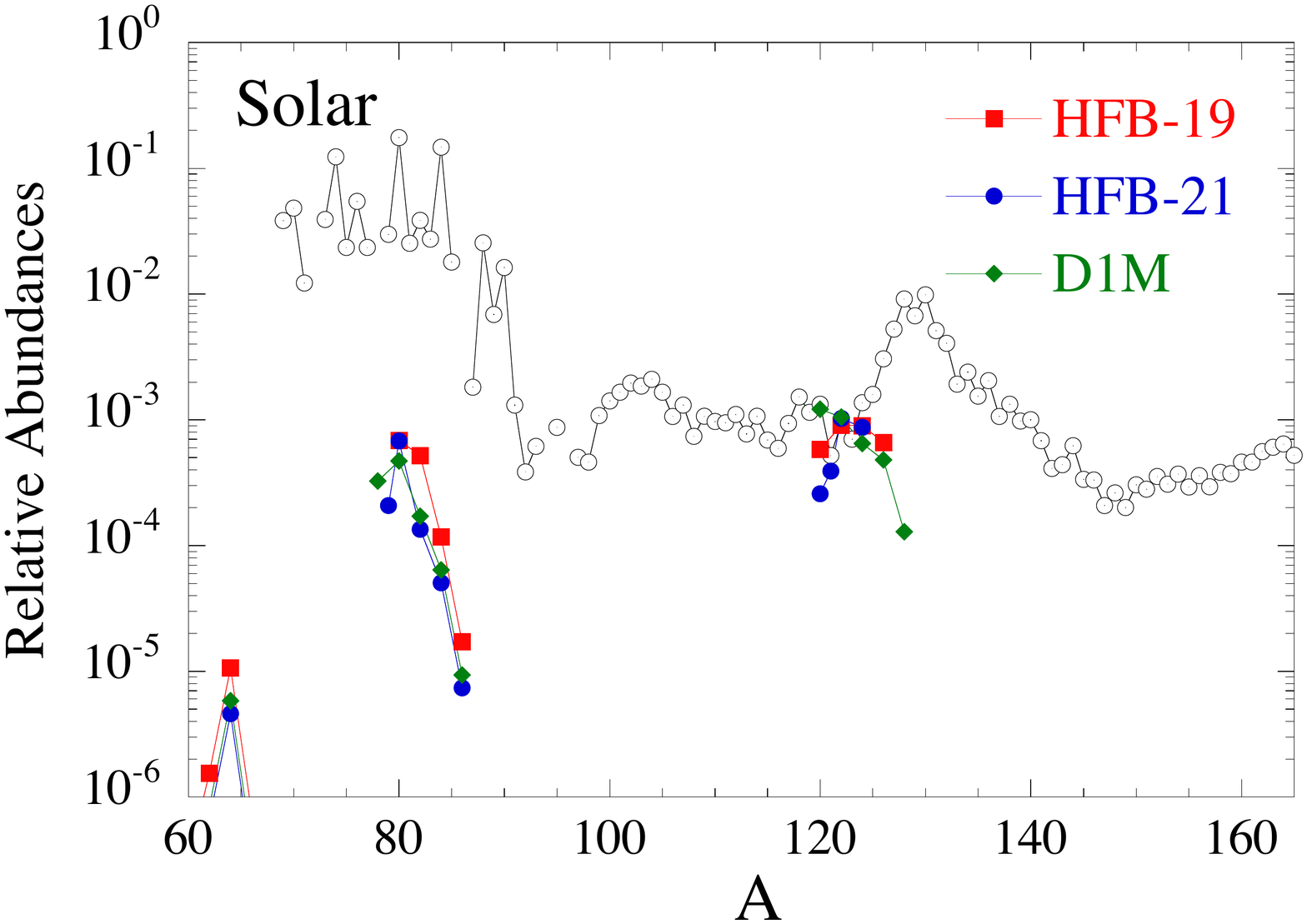}
\caption{Abundance distribution integrated over the whole outer crust for an initially cold NS of mass 1.5~\Msun and 13~km radius. The distributions are arbitrarily normalized to the solar system r-abundance distribution, shown for comparison. The distributions are obtained with the HFB-19 (squares), HFB-21 (circles) and D1M (diamonds) mass models when experimental masses (Audi et al. 2003, 2010) are not available. }
\label{fig02}
\end{figure}

The variation of the $T=0$ composition of the outer crust as a function of
density is illustrated in Fig.~\ref{fig01}. In this density range, the 
electron fraction $Y_e$ varies from 0.43 down to 0.31 at the drip density. 
Once in possession of the solution $\rho(r)$ to the TOV equations, 
it is easy to calculate the baryonic mass of any shell whose inner and outer
radii $r_1$ and $r_2$, respectively, are chosen to correspond to given
densities $\rho(r_1)$ and $\rho(r_2)$ (Pearson et al. 2011).  The abundances of the 
different nuclides in the outer crust can then be read off from 
Fig.~\ref{fig01}, the results being shown in Fig.~\ref{fig02}. Slightly 
different distributions are obtained with different mass models, but generally,
because of the high mass included within the high-density region (close to the
drip point), nuclides with $A\simeq 80$ or $118 \la A \la 126$ will be
dominant, as can be seen in Fig.~\ref{fig02}.

If a decompression of the $T=0$ outer crust material takes place, only $\beta$-decays towards the valley of stability are expected, no free neutrons being present (except through the possible $\beta$-delayed neutron emission). Consequently, after decompression only a restricted  distribution of $78 \la A \la 86$ and $120 \la A \la 126$ nuclei can be expected, as discussed in Sect.~\ref{sect_d0}. 

\subsection {The hot NS crust}
\label{sect_tt}

The $T=0$ picture of the NS crust is clearly an idealization, as it implies that the
star has had an infinite time to cool down and maintain (or restore) thermodynamic 
equilibrium since its creation in the aftermath of a gravitational-collapse 
supernova explosion. In reality, not only will the actual temperature of any 
particular NS be non-zero, but the equilibrium configuration 
corresponding to a still higher temperature might have become ``frozen in". The outer NS crust may in fact have a very different composition at non-zero temperature due to the specific softness of the distribution of the Gibbs free energy per nucleon. 

\begin{figure}
\centering
\includegraphics[scale=0.30]{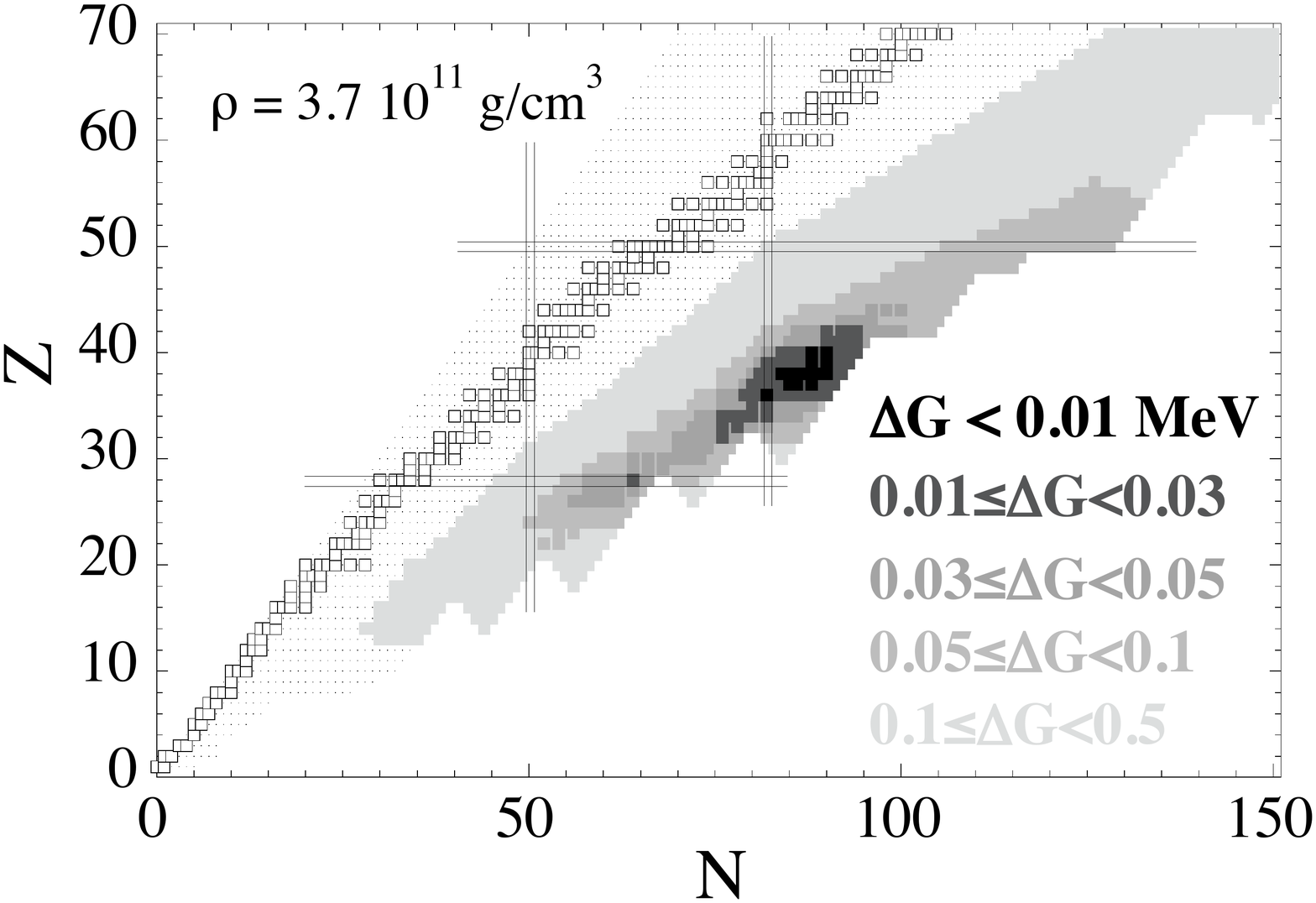}
\caption{Distribution of the Gibbs free energy per nucleon at $T=0$ and $\rho=3.7~10^{11}~{\rm g/cm^3}$. $\Delta G$ corresponds to the difference of the free energy per nucleon with respect to the minimum value obtained at this density for $Z=38$, $N=88$ and $A=124$. The distribution is obtained with the HFB-19 mass model (Goriely et al. 2010) when experimental masses are not available. The open squares represent stable nuclei.}
\label{fig03}
\end{figure}

As shown in Fig.~\ref{fig03}, many nuclei yield a free energy per nucleon that differs from the equilibrium one by no more than a few tens of keV. 
At finite temperatures, thermal fluctuations could therefore significantly broaden the distribution of equilibrium nuclides. At $T_9 \la 4-5$ (where $T_9$ is the temperature expressed in $10^9$K), an (n,$\gamma$)--($\gamma$,n) equilibrium may be established within the isotopic chain corresponding to the most probable $Z$ value at the given density. For higher temperatures,  an NSE could be reached and the abundance of a given nucleus $i$ of $Z$ protons and $N$ neutrons ($A=Z+N$) is now given by (Bravo \& Garcia-Senz 1999; Nadyoshin \& Yudin 2005; Arcones et al. 2010)
\be
Y_i=\frac{\omega_i}{\rho/m_u} \left(\frac{kTAm_u}{2\pi\hbar^2}\right)^{3/2} e^{N\eta_n+Z\eta_p}e^{Q_i/kT}
\label{eq_nse1}
\ee
where $\omega_i$ is the $T$-dependent partition function, and 
\be
Q_i=\left[ Zm_p+Nm_n-m_i\right]c^2+ Z\mu_{C,p} -\mu_C(Z,A) \ ,
\ee
includes the binding energy, expressed in term of the nucleon and nuclear masses $M_i$, as well as the Coulomb corrections to the chemical potential $\mu_C$ arising from the Coulomb contribution to the free energy, which becomes significant for heavier nuclei.
Eq.~\ref{eq_nse1} assumes that nuclei follow Maxwell-Boltzmann statistics, but the
presence of the degeneracy parameters $\eta_n$ and $\eta_p$ takes into account
the possibility that nucleons follow the more general Fermi-Dirac
statistics, which will certainly be the case for free neutrons close to the 
drip density. These parameters are related to the number density $n_q$ through 
the usual expression  
\be
n_q=\frac{8\pi\sqrt{2}}{h^3}m_q^3c^3\beta_q^3 \left[\mathcal{F}_{1/2}(\eta_q,\beta_q)+\beta_q\mathcal{F}_{3/2}(\eta_q,\beta_q) \right] \, ,
\ee
where $\beta=kT/mc^2$ is the relativistic parameter and $\mathcal{F}_k$ are the Fermi functions
\be
\mathcal{F}_k(\eta,\beta)=\int_0^{\infty}\frac{x^k(1+\beta x/2)^{1/2}}{e^{-\eta+x}+1} dx\, .
\ee
More details on the high-density NSE equations can be found in Bravo \& Garcia-Senz (1999) and Arcones et al. (2010)

To estimate the composition of the hot crust, the same EoS is used as in Pearson et al. (2011), though the $T$-dependent non-ideal Coulomb interaction between ions as well as between ions and electrons and between electrons in the approximation of a rigid electron background is taken from Haensel et al. (2007) and DeWitt et al. (1996). In this approximation, the Coulomb  chemical potential of species $i$ is given by 
\beqy
\mu_C(Z,A)=kT ~ f_C(\Gamma_i) 
\label{eq_mu}
\eeqy
where $f_C$ is the Coulomb free energy per ion in units of $kT$.  We would like to point out that we did not include here the contribution of $kT/3 \times \partial f_C / \partial {\rm ln}\Gamma_i$ as proposed by Glazyrin \& Blinnikov (2010), since this contribution implies a variation of the electron density and should therefore not be included in the NSE equations which are calculated at a fixed value of the electron density (see the discussion p.~4 of Glazyrin \& Blinnikov 2010). 

For a Coulomb liquid, $f_C$ can be expressed as (Haensel et al. 2007)
\beqy
f_C=& &\frac{F_C}{n_ikT}  =  A_1 \sqrt{\Gamma_i (A_2+ \Gamma_i)} \\ \nonumber
& & -  A_1 \times A_2 \ln\left( \sqrt{\Gamma_i/A_2}+\sqrt{1+\Gamma_i/A_2} \right) \\ \nonumber
& & +2 A_3\left[ \sqrt{\Gamma_i}-{\rm arctan}\left(\sqrt{\Gamma_i}\right)\right] \\ \nonumber
& & + B_1 \left[\Gamma_i-B_2 \ln\left(1+\frac{\Gamma_i}{B_2}\right) \right] +\frac{B_3}{2} \ln\left(1+\frac{\Gamma_i^2}{B_4}\right) \, ,
\eeqy
with $A_1=-0.9070$, $A_2=0.62954$, $A_3=0.27710$, $B_1=0.00456$, $B_2=211.6$, $B_3=-0.0001$ and $B_4=0.00462$. The corresponding contribution to the internal energy and the pressure can be found in Haensel et al. (2007).

\begin{figure}
\centering
\includegraphics[scale=0.30]{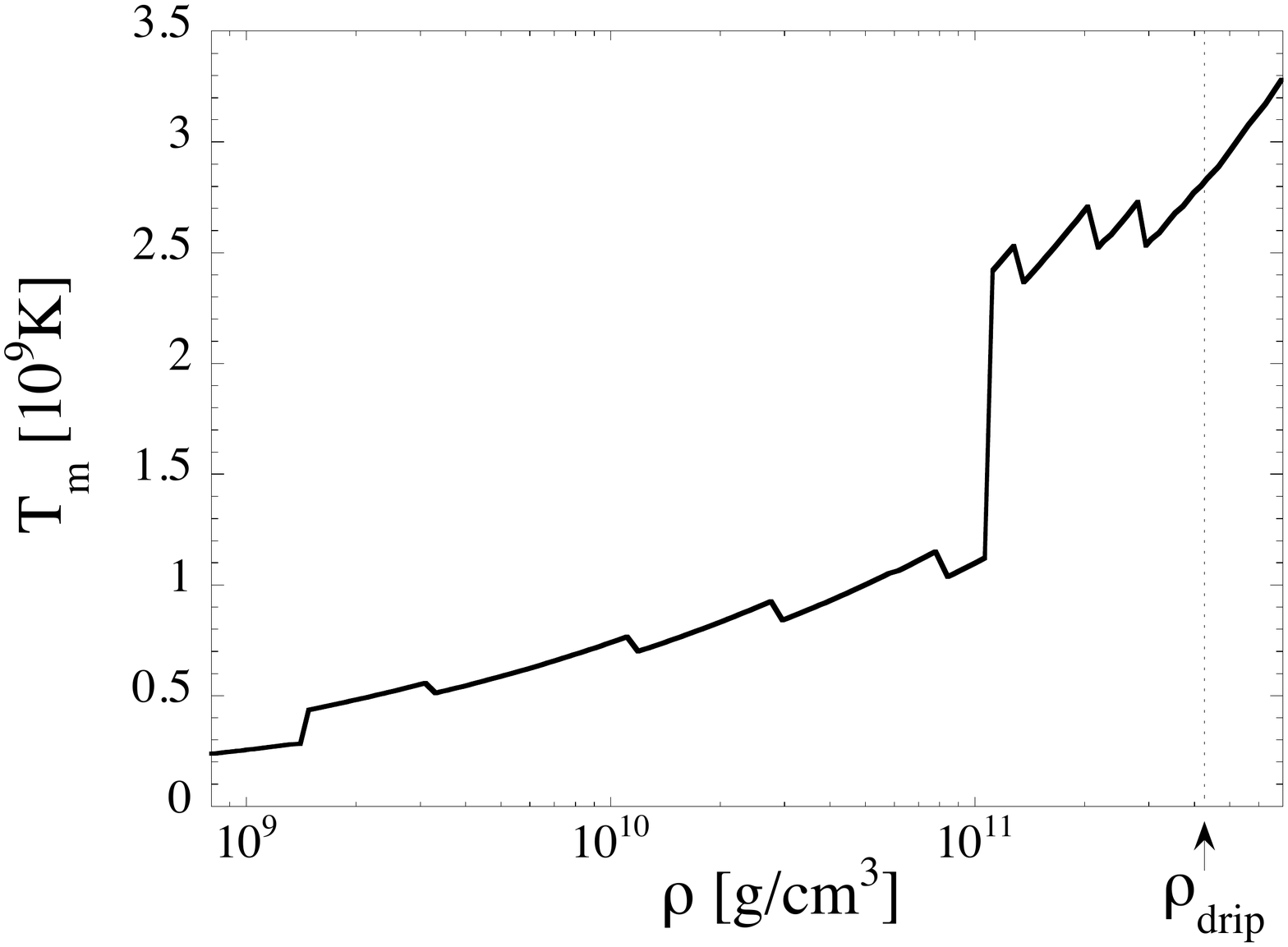}
\caption{Melting temperature $T_m$ as a function of the density $\rho$, estimated from Eq.~(\ref{eq_tm}) and the $T=0$ composition shown in Fig.~\ref{fig01} using HFB-19 masses when no experimental masses are available.}
\label{fig04}
\end{figure}

The Coulomb liquid approximation is adopted since at the temperatures considered here (typically $T_9=5-10$), the Coulomb coupling parameter,
\be
\Gamma_i=\frac{Z^2 e^2}{a_i kT} \, ,
\ee
where $a_i$ is the ion-sphere radius, remains smaller than the melting value $\Gamma_m = 175.0 \pm 0.4$ (Pothekin \& Chabrier 2000). Equivalently, the temperature $T$ is higher than the melting temperature,
\be
T_m=\frac{Z^2e^2}{a_ik \Gamma_m}\simeq 1.3~10^3~Z^2 \left(\frac{\rho{\rm [g/cm^3]}}{A}\right)^{1/3}~~ {\rm K} \, ,
\label{eq_tm}
\ee
as shown in Fig.~\ref{fig04}, where $T_m$ has been evaluated on the basis of the $T=0$ composition given in Fig.~\ref{fig01} (based on HFB-19 masses).

\begin{figure}
\centering
\includegraphics[scale=0.30]{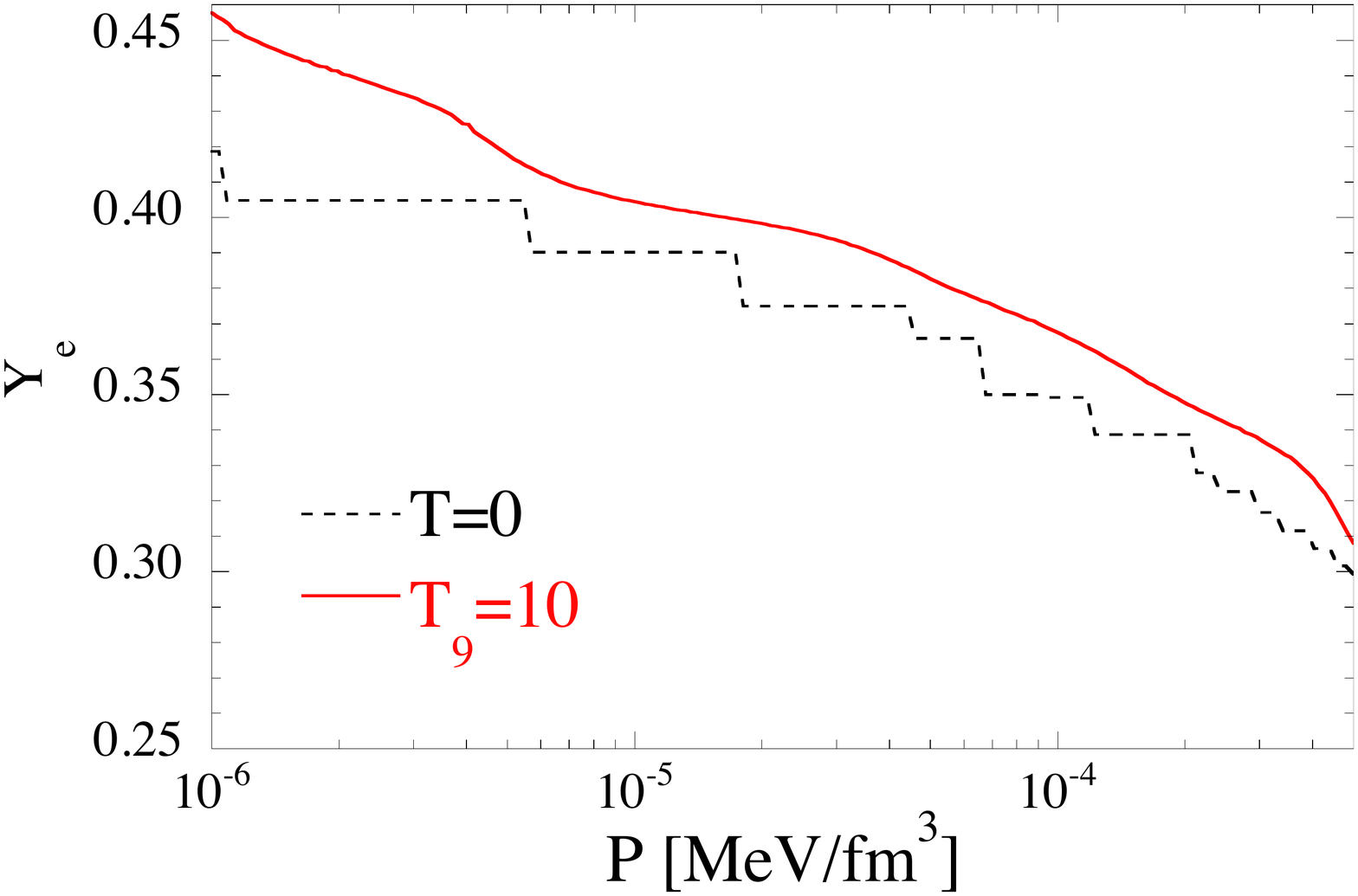}
\caption{$Y_e$ distribution in the outer crust, assuming (neutrino-less) $\beta$-equilibrium, as a function of the pressure. The dashed line corresponds to $T=0$ and the solid line to $T=10^{10}$~K.}
\label{fig05}
\end{figure}

To estimate the NSE composition of a hot NS crust still requires the 
determination of the distribution of the electron fraction $Y_e$, i.e., its
variation with the density $n$. We assume here that the NS cooled
sufficiently slowly for $\beta$-equilibrium to be maintained down to a 
temperature of about $T \simeq 10^{10}$~K, at which point the $Y_e$ 
distribution becomes frozen in as the NS continues to cool to lower
temperatures. However, we assume that NSE is maintained down to temperatures of
the order of $T \simeq 7-10~10^{9}$~K (see Sect.~\ref{sect_dt} for a discussion
of the NSE timescales). The $Y_e$ distribution corresponding to 
$\beta$-equilibrium is obtained by minimizing the free Gibbs energy per nucleon at $T =10^{10}$~K;
for the physics used, see also Pearson et al. (2011). We use the traditional 
linear mixing rule (also known as the additive approximation) to estimate the 
total free Gibbs energy of the multicomponent plasma by weighting the 
individual free energy of each species by its molar fraction in the mixture. 
With respect to the $T=0$ case, the $Y_e$ distribution is found to be slightly 
higher, and obviously much smoother, as shown in Fig.~\ref{fig05}. The implied
decrease in the neutron fraction occurs despite the liberation of free neutrons
at non-zero temperatures.

\begin{figure}
\centering
\includegraphics[scale=0.30]{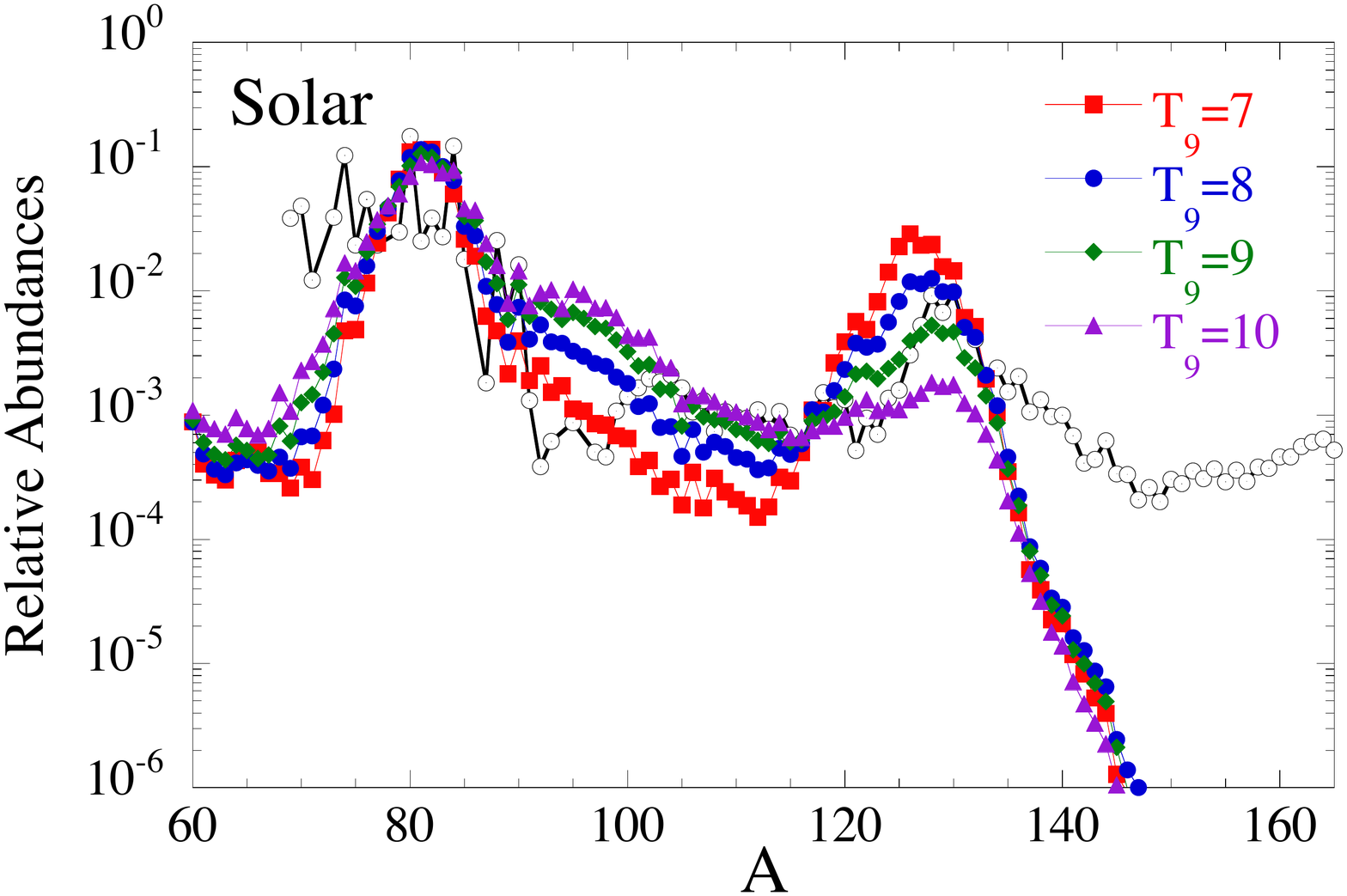}
\caption{NSE abundance distribution integrated over the whole outer crust for 4 different temperatures assuming the initial $Y_e$ distribution is determined by $\beta$-equilibrium at $T=10^{10}$~K.
The distribution is obtained with the HFB-19 mass model (Goriely et al. 2010) when experimental masses are not available. The solar system r-abundance distribution (dotted circles) is shown for comparison.}
\label{fig06}
\end{figure}

\begin{figure}[htbp]
\centering
\includegraphics[scale=0.30]{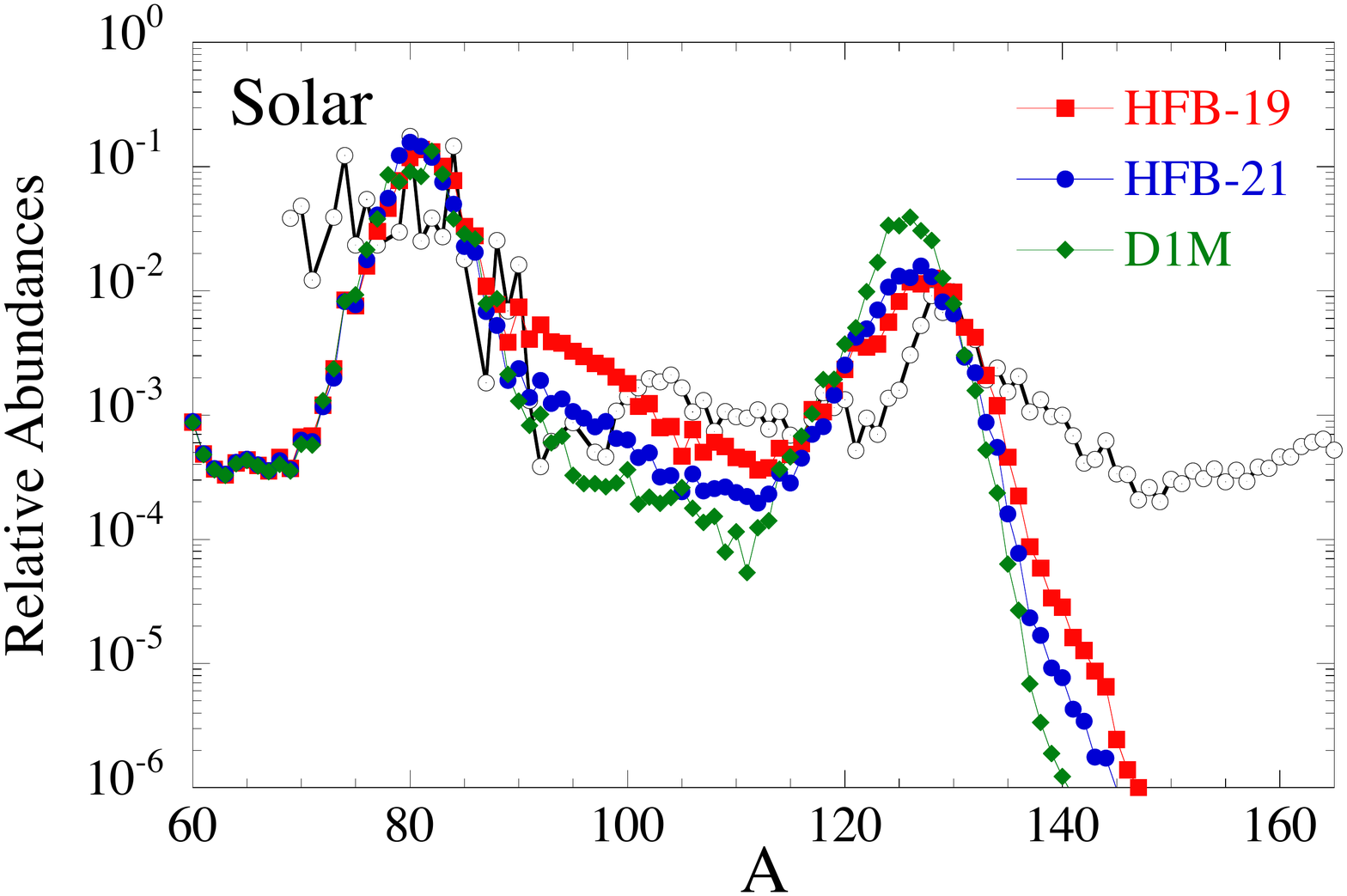}
\caption{Same as Fig.~\ref{fig06} for $T_9=8$ and different mass models, namely HFB-19, HFB-21 (Goriely et al. 2010) or D1M (Goriely et al. 2009).}
\label{fig07}
\end{figure}

\begin{figure}[htbp]
\centering
\includegraphics[scale=0.30]{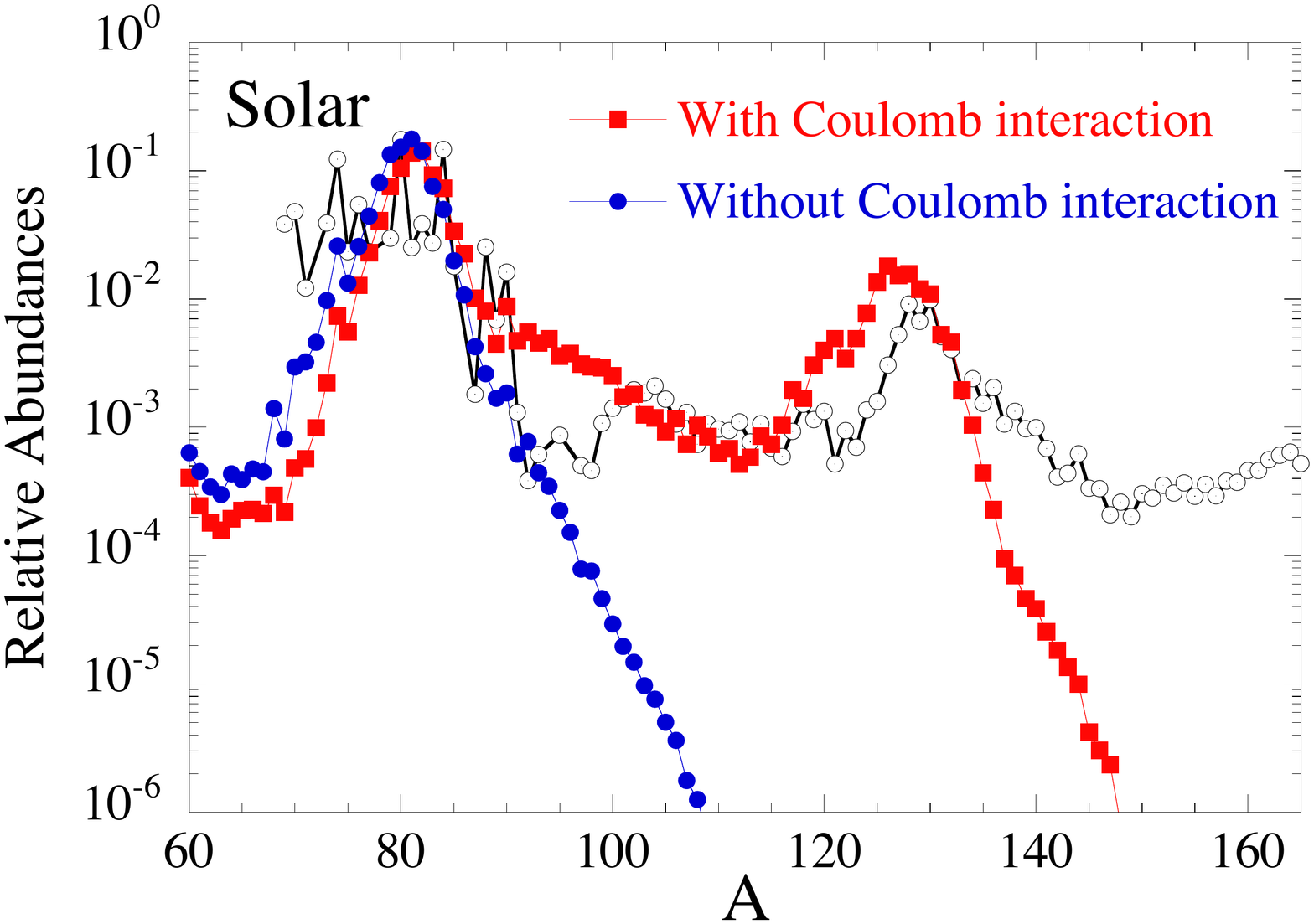}
\caption{Same as Fig.~\ref{fig07} including and excluding the Coulomb interaction (the HFB-19 mass model is used).}
\label{fig08}
\end{figure}

The NSE abundance distributions for the whole outer crust at four different  temperatures (between 7 and 10~GK) and integrated over a pressure  column given by the TOV equations are shown in Fig.~\ref{fig06}.    
The outer NS crust in NSE at a temperature of $T_9 \simeq 7-10$ is seen to be 
made of r-nuclei with a distribution similar to that of the solar system, 
although the second peak  is slightly shifted to lower masses around $A\simeq 126$, as is also found for the $T=0$ composition  (Fig.~\ref{fig02}). The  second
peak is also found to be rather pronounced with respect to the first peak for 
low temperatures ($T_9=7$), but to be smoothed away, on the other hand, for 
high temperatures ($T_9=10$). The abundance  distribution is sensitive to the 
adopted mass models, as illustrated in Fig.~\ref{fig07}, but is otherwise 
extremely robust, since as far as the nuclear physics is concerned it depends 
only on binding energies. The Coulomb correction plays a vital role by shifting
significantly the abundance distribution towards the high-mass region.
In particular, no $N \simeq 82$ would be found in the crust if the  Coulomb interaction was omitted in the NSE equations, as shown in Fig.~\ref{fig08}.

\section{Decompression of the NS outer crust}
\label{sect_d}

Without focussing clearly on a particular ejection scenario of outer-crust 
material of an NS, we decribe the density evolution of an ejecta clump
in simple, but as general terms as possible. For setting the 
values of the free parameters 
we will refer to existing results from simulations of the dynamical merging 
of binary NSs or an NS with a black hole. In these events
near-surface matter can be tidally stripped and ejected, and we take the
ejection dynamics of such events as template for our expansion model.

The evolution of the matter density is approximated by a one-zone model
of the pressure-driven expansion of a self-gravitating clump under the
influence of tidal forces (exerted by the compact source object of
the ejecta) along the escape 
trajectory of the clump. In order to determine the time evolution of the 
clump radius $R(t)$, we integrate numerically the Newtonian equation of 
motion 
\begin{equation}
\frac{\mathrm{d}^2 R}{\mathrm{d}t^2} = -\frac{4\pi}{3}G\rho R + 
\frac{P}{\rho R} + \frac{R}{\left (\tau_{\mathrm{esc}}+\frac{3}{2}t 
\right)^2} \,.
\label{eq_acc}
\end{equation}
Here the first term corresponds to the confining acceleration by the
self-gravity of the clump, approximated by $-GM_{\mathrm{c}}/R^2\approx
-\frac{4\pi}{3}G\rho R$ when $M_{\mathrm{c}} 
\approx \frac{4\pi}{3}\rho R^3$ is the
clump mass with average density $\rho$. The inflating effect of 
the gas pressure $P$ of the clump is repesented by the second term in 
Eq.~(\ref{eq_acc}), where we use a one-zone representation of the 
pressure gradient according to $\rho^{-1}\mathrm{d}P/\mathrm{d}r \approx
P/(\rho R)$. The third term describes the acceleration corresponding
to the tidal stretching by the external gravitational force of the NS,
\begin{equation}
\left ( \frac{\mathrm{d}^2 R}{\mathrm{d}t^2}\right )_{\mathrm{tidal}}
\sim \frac{GM_{\mathrm{ns}}}{r_\mathrm{c}^2} - 
\frac{GM_{\mathrm{ns}}}{(r_\mathrm{c}+R)^2}
\approx \frac{2 GM_{\mathrm{ns}}R}{r_\mathrm{c}^3} \,,
\label{eq_tidal}
\end{equation}
where $M_{\mathrm{ns}}$ is the NS mass and $r_\mathrm{c}(t)$ the 
time-dependent radial position of the clump center. In the transformation
leading to the last term in Eq.~(\ref{eq_tidal}) we made use of the 
relation $R \ll r_\mathrm{c}$. Assuming that the
ejection velocity $v_\mathrm{c} = \mathrm{d}r_\mathrm{c}/\mathrm{d}t$ 
of the clump is given by the escape velocity on a parabolic orbit, i.e.,
$\frac{1}{2}v_\mathrm{c}^2 = GM_{\mathrm{ns}}/r_\mathrm{c}$, we can 
solve the equation of motion of the clump center,
$\mathrm{d}r_\mathrm{c}/\mathrm{d}t = \sqrt{2GM_{\mathrm{ns}}/r_\mathrm{c}}$,
to obtain 
\begin{equation}
r_\mathrm{c}(t) = r_0\left (1 + \frac{3}{2}\frac{t}{\tau_{\mathrm{esc}}}
\right )^{2/3} \,,
\label{eq_rc}
\end{equation}
with $r_0$ being the radial position of the clump center at the 
beginning of the ejection, $\tau_{\mathrm{esc}} = r_0/v_{\mathrm{esc}}$
the escape timescale, and $v_{\mathrm{esc}} =
\sqrt{2GM_{\mathrm{ns}}/r_0}$ the escape velocity. Using 
$r_\mathrm{c}(t)$ of Eq.~(\ref{eq_rc}) in Eq.~(\ref{eq_tidal}) we get
\begin{equation}
\left ( \frac{\mathrm{d}^2 R}{\mathrm{d}t^2}\right )_{\mathrm{tidal}} 
\sim \frac{R}{\left (\tau_{\mathrm{esc}} + \frac{3}{2}t \right )^2} \,,
\label{eq_tidac}
\end{equation}
which appears as the last term on the rhs of Eq.~(\ref{eq_acc}).
Analytic estimates for the conditions at the surface of an NS in
agreement with numerical results of NS merger simulations 
(Ruffert \& Janka 2001) yield values of the escape timescale
$\tau_{\mathrm{esc}}$ in the range of $10^{-4}$ to $3~10^{-4}$~s.
The value of $\tau_{esc} =3~10^{-4}$~s is adopted in this work.

In solving the simple model equation, Eq.~(\ref{eq_acc}), the initial
clump radius $R_0$ plays an important role, because it determines
the initial acceleration of the clump expansion and therefore its 
expansion timescale. It can be used as a free parameter, leading to
different density decline rates during decompression.
Here we choose a value of $R_0 \approx 2\,$km,
which is found to reproduce the density evolution of the majority 
of NS merger ejecta (Ruffert \& Janka 2001) fairly well. The pressure
$P$ of the clump medium is determined from the EoS of Timmes \& Arnett 
(1999) and is consistently evolved for the changing conditions of
density and temperature (including possible $\beta$-decay heating)
during the decompression history. The 
integration of Eq.~(\ref{eq_acc}) yields the time-dependent clump
radius $R(t)$, with which (for constant clump mass) the clump
density can be computed as $\rho(t) = \rho_0(R_0/R(t))^3$.

The composition change during the decompression is followed with a full network, where in particular the $\beta$-decay processes have been taken from the updated version of the Gross Theory (Tachibana et al. 1990)  and neutron capture rates are consistently estimated with the TALYS code (Goriely et al. 2008) on the basis of the nuclear mass model used for the initial conditions. The temperature 
evolution is followed as described in Meyer (1989) and Goriely et al. (2005) on
the basis of the laws of thermodynamics and possible nuclear heating through 
$\beta$-decay processes. The decompression along the above-defined trajectory of both the cold NS outer crust and the outer crust initially in NSE are studied below.

\subsection{The cold NS crust}
\label{sect_d0}

\begin{figure}
\centering
\includegraphics[scale=0.31]{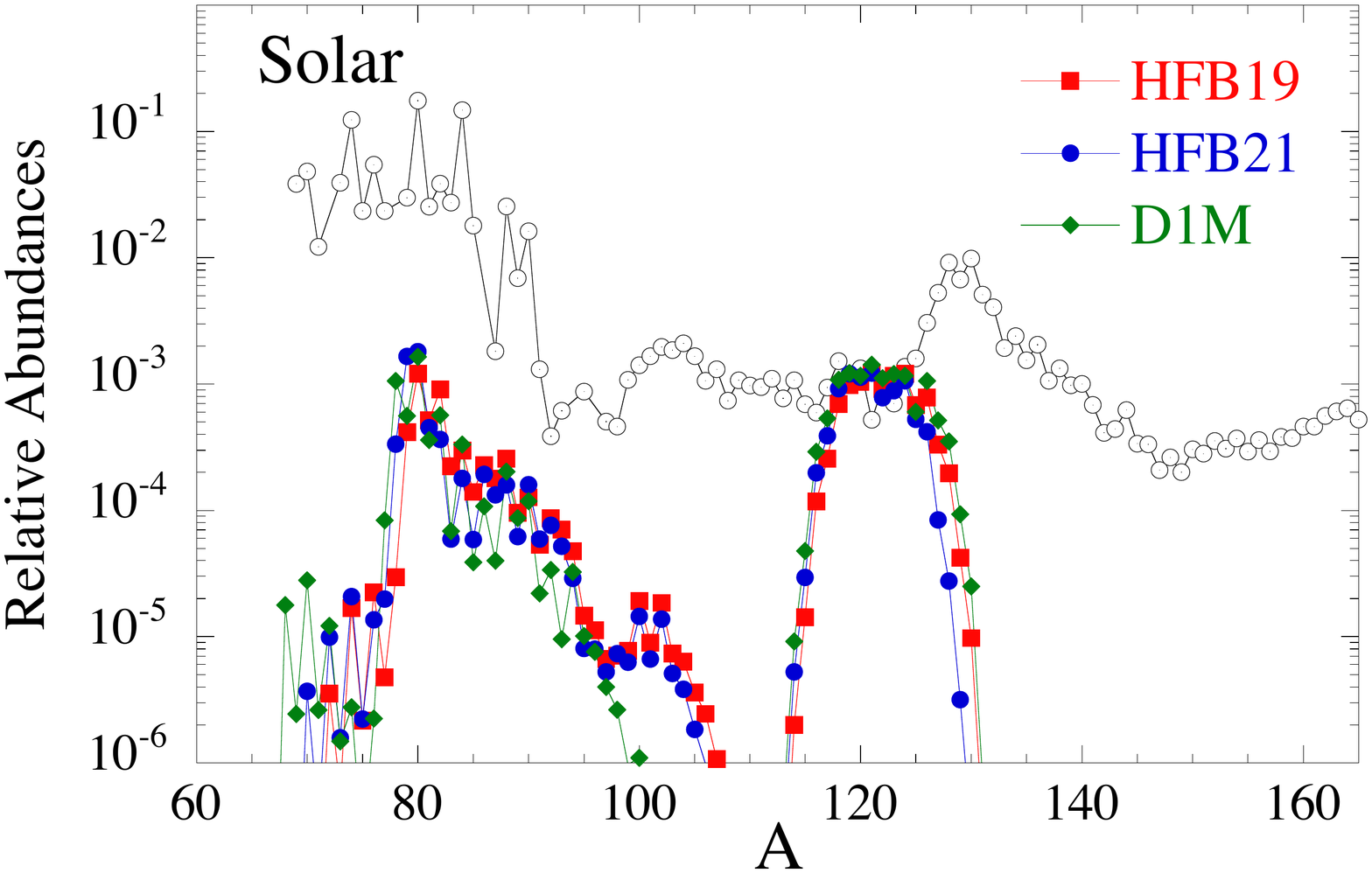}
\caption{Same as Fig.~\ref{fig02} {\it after decompression}. }
\label{fig09}
\end{figure}

The final isobaric abundance distribution of the matter ejected from the cold NS crust initially assumed to be at $\beta$-equilibrium can be expected to  remain relatively similar to the one prior to the ejection, since only $\beta$-decay and $\beta$-delayed neutron emission (with the possibility of recapturing the emitted neutrons) may change the initial composition.
The final isobaric abundance distribution is shown in Fig.~\ref{fig09}, the initial composition being given in Fig.~\ref{fig02} prior to the decompression. Globally, the final distribution is obviously far from matching the overall solar system pattern. Some differences are seen depending on the nuclear physics ingredients adopted. Interestingly, the decompression of the cold NS outer crust appears as a potential site for producing almost exclusively the $115 \la A \la 124$ r-nuclei. In some  site-independent parametrized r-process models, these nuclei have been found to be underproduced, though this conclusion is obviously subject to large nuclear physics and astrophysics uncertainties (for a review, see
Arnould et al. 2007). 

\subsection{The hot NS crust}
\label{sect_dt}
\begin{figure}
\includegraphics[scale=0.30]{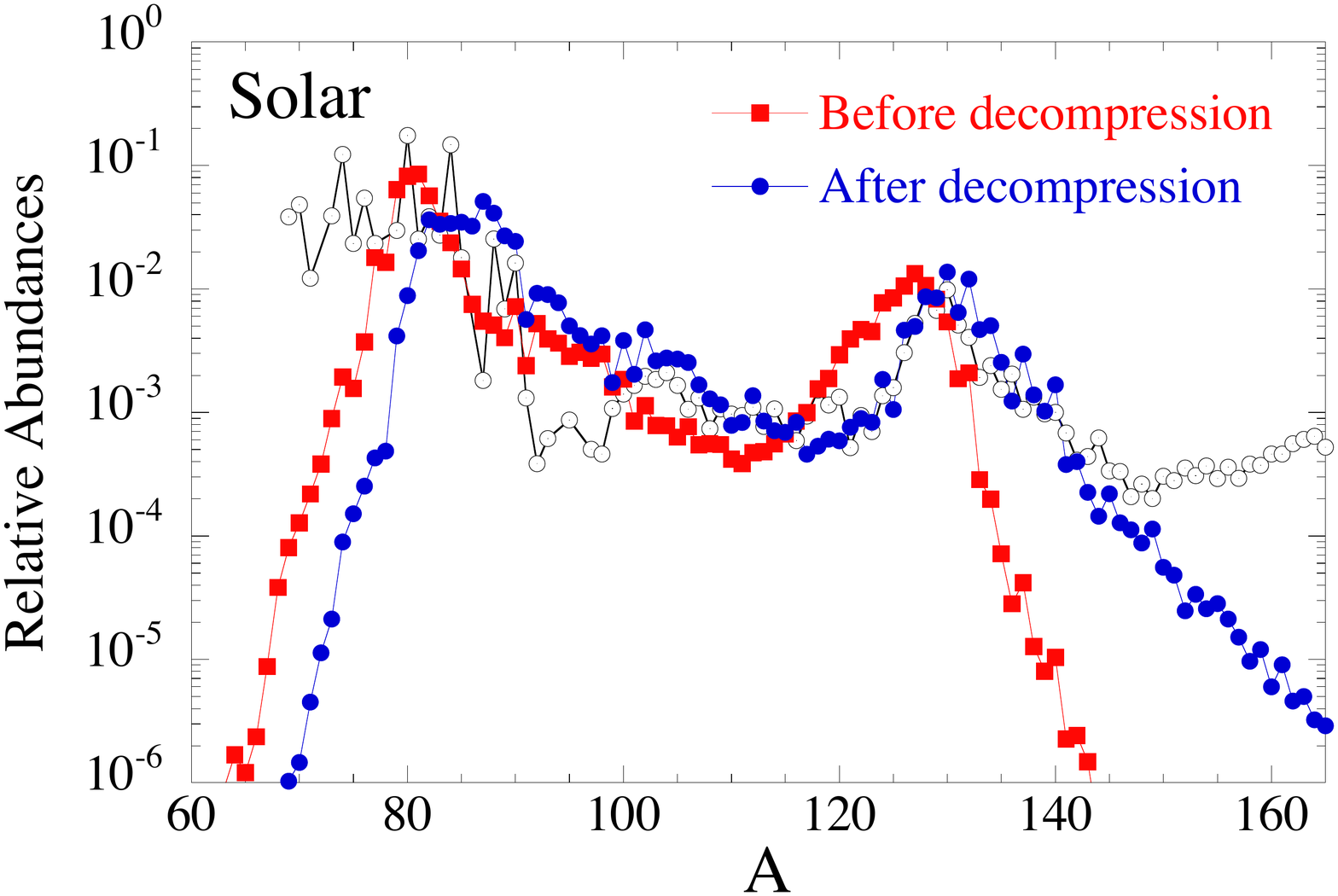}
\caption{Abundance distribution before and after decompression of a layer with initial pressure $P_0=4~10^{-4}$~MeV/fm$^3$ and density $\rho_0=3.4~10^{11}{\rm g/cm^3}$ and initially in NSE at $T_9=8$.  The calculation was performed with the HFB-19 masses and corresponding reaction rates.}
\label{fig10}
\end{figure}

\begin{figure}
\includegraphics[scale=0.30]{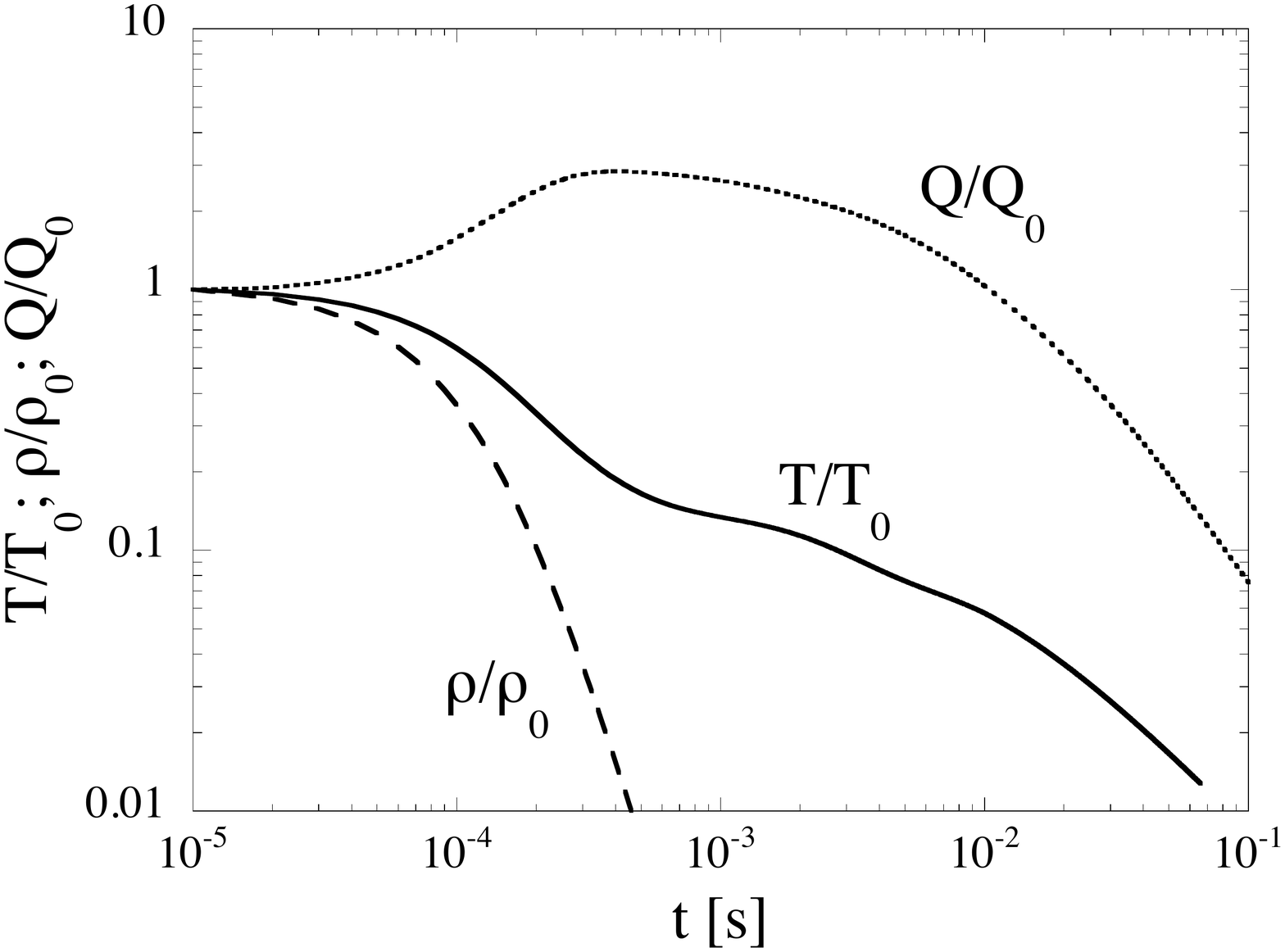}
\caption{Evolution of the temperature, density and radioactive heating rate per unit mass ($Q$) resulting from the decompression of the layer with an initial pressure $P_0=4~10^{-4}$~MeV/fm$^3$ and density $\rho_0=3.4~10^{11}{\rm g/cm^3}$ and initially in NSE at $T_0=8~10^9$~K. The initial radioactive decay heat $Q_0$ amounts to $6.6~10^{18}$~erg~g$^{-1}$~s$^{-1}$.}
\label{fig11}
\end{figure}

As discussed in Sect.~\ref{sect_tt}, the outer crust initially in NSE
at temperatures around $T_9=7-10$ differs considerably from the cold $T=0$ case. In addition, for layers close to the drip density, free neutrons are present in significant numbers due to the high initial temperatures. In this case, a few  neutrons per seed nucleus are  available and may be captured during the  expansion phase so that the initial NSE mass distribution is modified. The abundance distributions  in a layer with initial pressure $P_0=3~10^{-4}$~MeV/fm$^3$ and density $\rho_0=2.7~10^{11}{\rm g/cm^3}$, initially in NSE at $T_9=8$ is shown in Fig.~\ref{fig10}. The evolution of the temperature,  density  as well as the corresponding radioactive power due to  $\beta$-decays is shown in Fig.~\ref{fig11} for the same ejected layer as the one considered in Fig.~\ref{fig10}. In particular, it can be seen that the $\beta$-decay heating slows down the temperature drop already a few ms after ejection.

After decompression, due to the capture of about 5 free neutrons per seed nucleus, the distribution is shifted towards the $A\simeq 130$ peak and shaped by the lower temperatures found at the time of the neutron captures.
The final distribution is found to be independent of the details characterizing the expansion. In particular for a faster expansion obtained with the value of $\tau_{\mathrm{esc}}=10^{-4}$~s a virtually identical distribution is found.

\begin{figure*}
\centering
\includegraphics[scale=0.50]{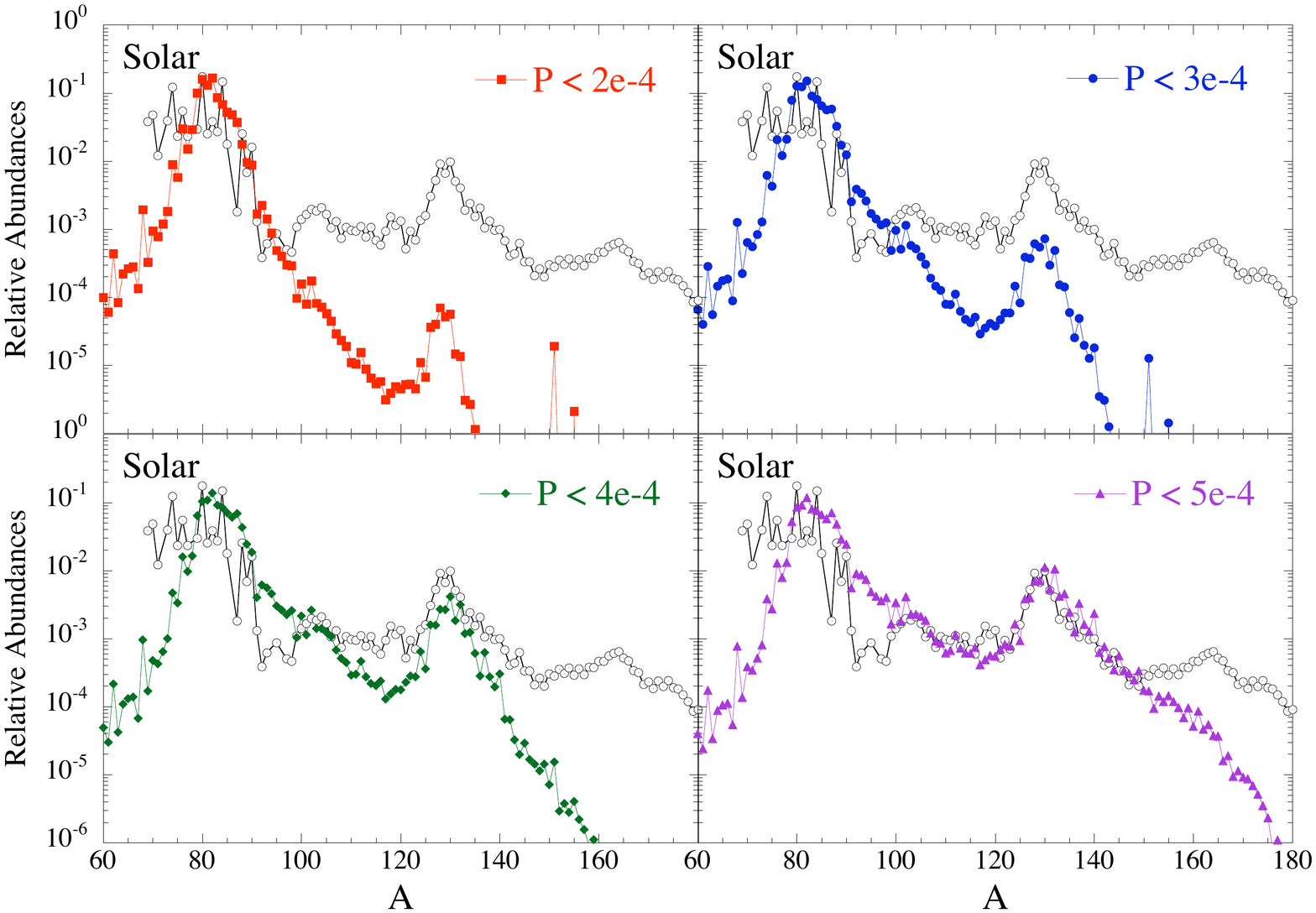}
\caption{Final abundance distributions for the whole outer crust (as integrated up to a pressure $P$ given in the legend in MeV/fm$^3$) after decompression when initially in NSE at $T_9=8$. The calculations are performed with the HFB-19 mass model and corresponding rates. The distributions are compared with the solar r-abundance distributions (dotted circles).}
\label{fig12}
\end{figure*}

The abundance pattern for  the outer crust  integrated up to a maximum pressure $P$ is given in Fig.~\ref{fig12} for the initial temperature of $T_9=8$. If the whole outer crust is considered ($P\le P_{\rm drip}=5~10^{-4}$~MeV/fm$^3$), an overall agreement with the solar distribution is obtained in the whole  $80 \la A \la 150$ mass region. An overproduction of the $A>140$ r-nuclei (relative to the solar distribution) could be obtained if a sizable part of the inner crust was ejected at the same time (Arnould et al. 2007). When the mass ejection is restricted to parts of the outer crust (up to a pressure value $P<P_{\rm drip}$), different r-abundance distributions are obtained, as shown in Fig.~\ref{fig12}. The production of r-nuclei in such a scenario is clearly sensitive to the thickness of the outer crust that is ejected and therefore depends on the ejection mechanism that is invoked.

\begin{figure*}
\centering
\includegraphics[scale=0.50]{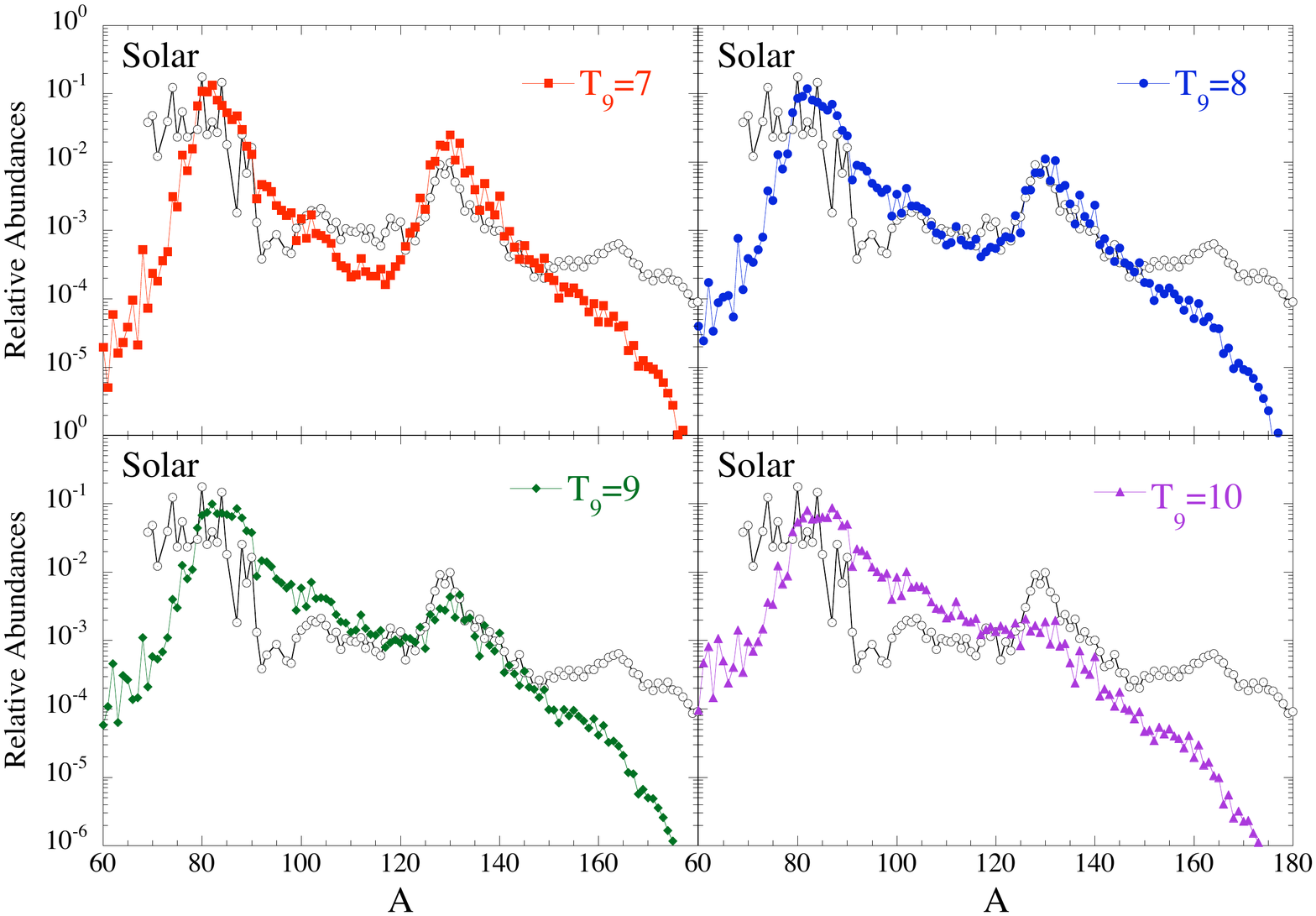}
\caption{Final abundance distributions of the outer crust material after decompression, if initially in NSE at different temperatures ranging between $T_9=7$ and $T_9=10$. The calculations are performed with the HFB-19 mass model and corresponding rates. The distributions are compared with the solar r-abundance distributions (dotted circles).}
\label{fig13}
\end{figure*}

The overall abundance distribution depends not only on the fraction of the ejected crust but also on the initial temperature at which the NSE has been frozen in. We show in Fig.~\ref{fig13} the abundance distributions obtained for different initial temperatures between 7 and 10~$10^9$~K,  assuming that the whole outer crust is ejected. 

\begin{figure}
\includegraphics[scale=0.31]{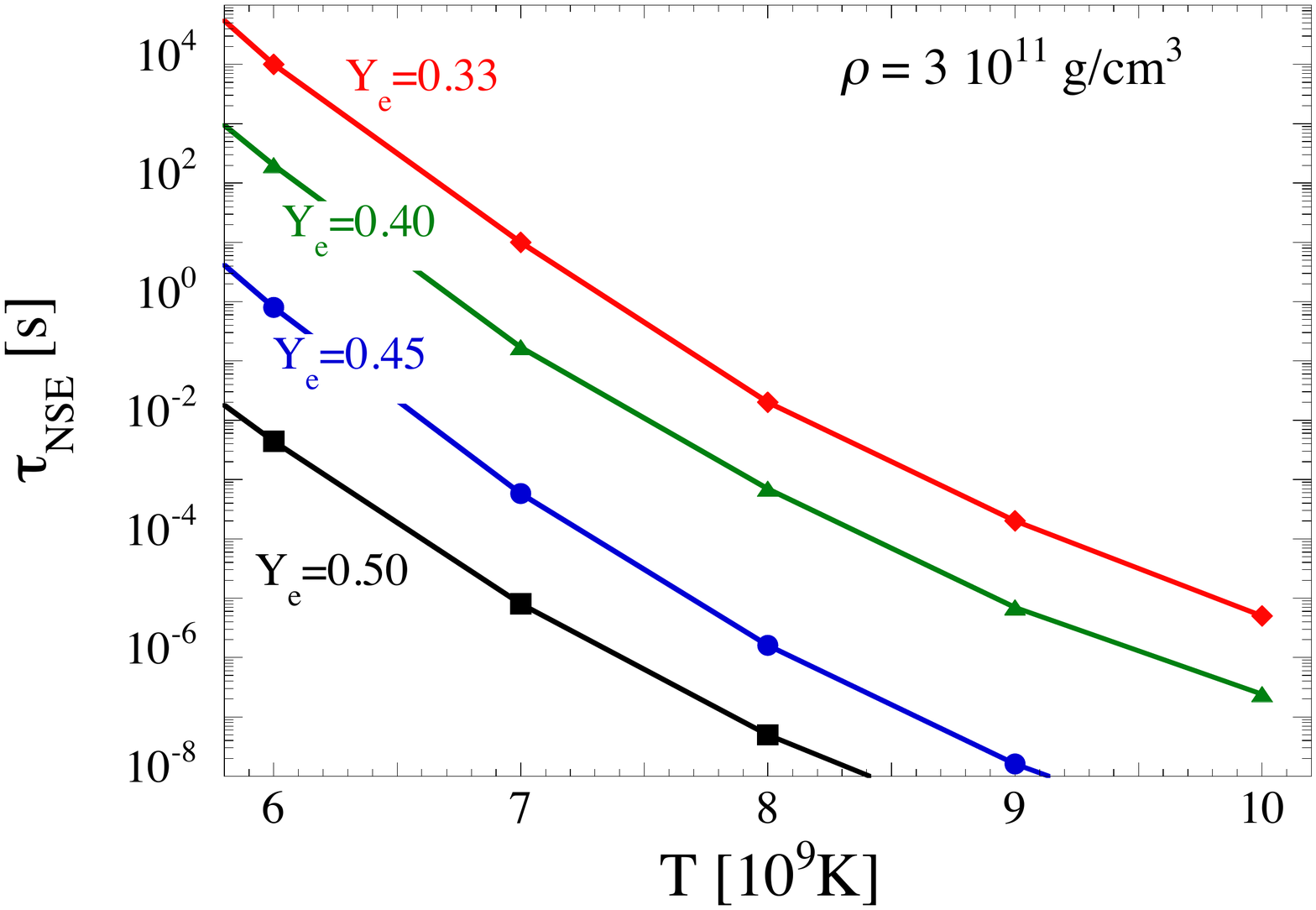}
\caption{Time needed to reach NSE as a function of the temperature for different values of $Y_e$ at a constant density of $\rho=3~10^{11}~{\rm g/cm^{-3}}$. Symbols corresponding to the same value of $Y_e$ are connected by a solid line. $Y_e=0.50$ is shown by squares, 0.45 by circles, 0.40 by triangles and 0.33 by diamonds.}
\label{fig14}
\end{figure}

It could be seen as purely fortuitous that temperatures around $T_9\simeq 8$ give rise to r-abundance distributions in agreement with the solar distribution. However this is in agreement with the astrophysical scenario considered here.
In particular,  the temperatures considered here typically correspond to those at which NSE can be dynamically achieved in cooling events. We have calculated the typical timescale $\tau_{NSE}$ needed to reach NSE for a density of $\rho=3~10^{11}~{\rm g/cm^{-3}}$ (typical of the most contributing part of the outer crust) and for different values of $Y_e$ (as found in the outer crust). This timescale is obtained by estimating the time needed to reach a steady state at constant temperature and density considering different initial seed nuclei for the chosen value of $Y_e$. A full network including all strong and electromagnetic interactions (but no weak interactions) is used for this purpose. As shown in Fig.~\ref{fig14}, for matter with initial values of $Y_e=0.33-0.40$, it takes about 1 to 20~ms to reach an NSE at $T_9=8$ and $\rho=3~10^{11}~{\rm g/cm^{-3}}$, while at $T_9\simeq 9$, around 0.2~ms are required for the most neutron-rich conditions ($Y_e=0.33$). Such ms timescales are characteristic of dynamical scenarios of interest here for the potential mass ejection (e.g.\ in the bursts of soft gamma-repeaters, during NS mergers, \dots), so that prior to the ejection the NSE should be achieved at temperatures typically above $T_9\simeq 8-9$ and can be expected to be frozen below during the ejection.  
The final production of r-nuclei is clearly sensitive to the temperatures at which NSE freezes in so that depending on the dynamical timescales different distributions (as seen in Fig.~\ref{fig13}) are found.

\begin{figure}
\includegraphics[scale=0.41]{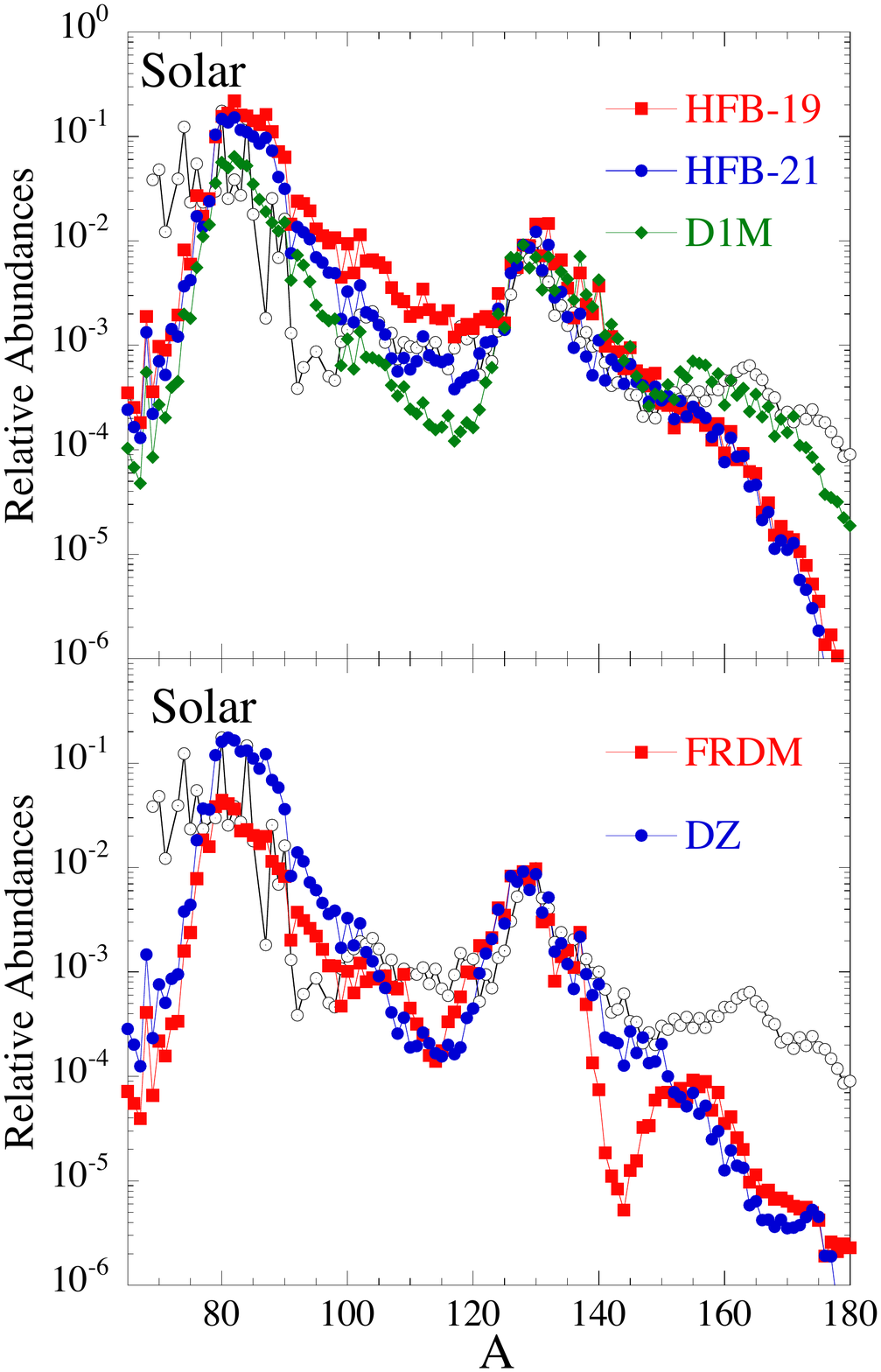}
\caption{Same as Fig.~\ref{fig13} for an initial NSE temperature of  $T_9=8$ and five different mass models (and corresponding reaction rates), namely the HFB-19, HFB-21 and D1M mass models (upper panel) and the Duflo \& Zuker (DZ; 1995) and FRDM (M\"oller et al. 1995) models (lower panel).}
\label{fig15}
\end{figure}

\begin{figure}
\includegraphics[scale=0.30]{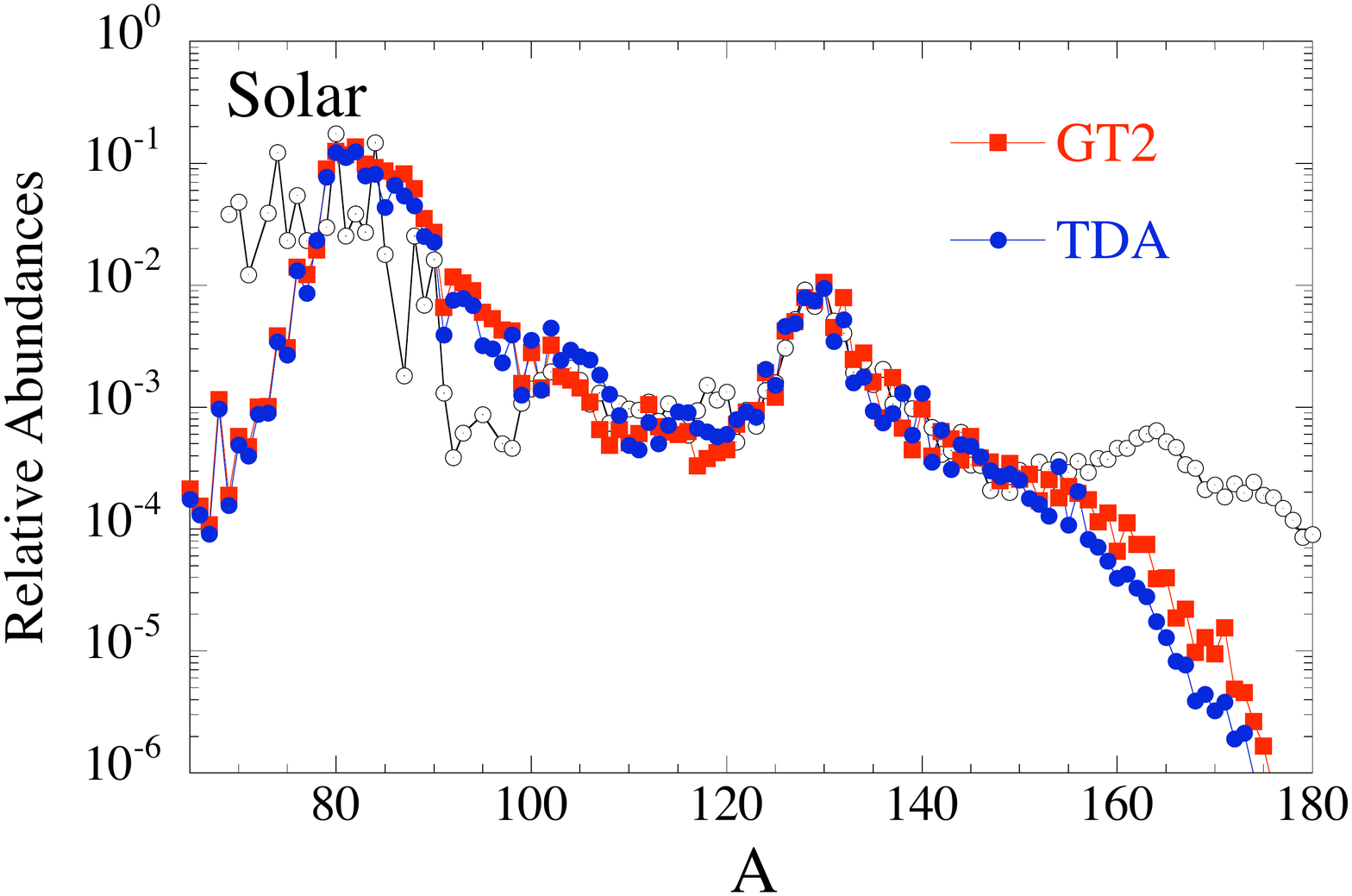}
\caption{Same as Fig.~\ref{fig13} for an initial NSE temperature of  $T_9=8.5$ and two different calculations of the $\beta$-decay rates, namely the Gross Theory (version 2; Tachibana et al. 1990) and the Tamm-Dancoff approximation (TDA) of Klapdor et al. (1984). The calculations were performed with the HFB-21 mass model and corresponding rates.}
\label{fig16}
\end{figure}

Uncertainties in the nuclear physics input can influence the calculation of the final abundances. These concern essentially nuclear masses and the corresponding reaction rates. We show in Fig.~\ref{fig15} the abundance distributions obtained with NSE distributions and reaction rates determined on the basis of different mass models. The initial NSE temperature is 8~$10^9$~K and the whole outer crust mass is assumed to be ejected. As shown in Fig.~\ref{fig15}, the exact strength, location, and width of the $N=50$ and $N=82$ r-process peaks are sensitive to the mass model adopted. The production of the $90 \la A \la 120$ r-nuclei can also be significantly modified according to the nuclear physics input. The importance of nuclear masses in this specific r-process nucleosynthesis scenario is essentially linked to the initial NSE conditions, which define not only the most abundant species initially present in the outer crust, but also the amount of free neutrons available for the neutron-capture process during the decompression. In contrast, the impact of uncertainties affecting the $\beta$-decay rates is found to be rather limited (as illustrated in Fig.~\ref{fig16}), though $\beta$-delayed neutron emission plays some role in smoothing the final abundance pattern.

\section{Conclusion}

Most of the problems faced in understanding the origin of r-process elements
and observed r-abundances are related to our ignorance of the astrophysical site that is
capable of providing the required  high neutron flux. We have shown here that the decompression of 
the NS matter from the outer crust provides suitable conditions for a
robust r-processing of the light species, i.e., r-nuclei with $A\le 140$, 
provided the outer crust is initially in NSE at temperatures around 
$T_9\simeq 8$ and the ejection mechanism allows for the full outer crust to be expelled to the interstellar medium. During the decompression, the free neutrons (initially liberated by the high temperatures) are re-captured leading to a final pattern similar to the solar system distribution.
 
While the total mass included in the outer crust only amounts to some $10^{-5}-10^{-4}$~\Msun, its ejection leads to a galactic enrichment in stable neutron-rich nuclei with a final composition close to what is now observed in the solar system. The final composition should carry the imprint of the initial temperature at which the NSE is frozen in prior to the ejection, as well as the density region of the outer crust that gets ejected. The final abundances are also affected by nuclear uncertainties, most particularly nuclear masses. The similarity between the predicted and solar abundance patterns as well as the robustness of the prediction against variations of input parameters (such as expansion timescales or initial $Y_e$ distribution) make this site one of the most promising that has been proposed, deserving further  exploration with respect to various aspects such as nucleosynthesis,  hydrodynamics, and galactic chemical evolution.

\begin{acknowledgements}
The authors are thankful to D.G. Yakovlev and A.Y. Potekhin for valuable discussions. S.G. and N.C acknowledge the financial support of the "Actions de recherche concert\'ees (ARC)" from the "Communaut\'e fran\c caise de Belgique". S.G and N.C are F.N.R.S. research associates. HTJ acknowledges support by the Deutsche Forschungsgemeinschaft through the Transregional Collaborative Research Centers SFB/TR~27
``Neutrinos and Beyond'' and SFB/TR~7 ``Gravitational Wave Astronomy'', and the Cluster of Excellence EXC~153 ``Origin and Structure of the Universe'' ({\tt http://www.universe-cluster.de}).
\end{acknowledgements}

\end{document}